\documentclass[aps,prb]{revtex4}
\usepackage{amsmath}
\usepackage{graphicx}

\begin{document}

\title{Anomalous Doppler Effect Singularities in    Radiative 
Heat Generation, Interaction Forces  and Frictional Torque 
for  two Rotating Nanoparticles}

\author{A.I.Volokitin$^{1,2*}$ }

\affiliation{
$^1$Samara State Technical University, Physical Department, 443100 Samara, Russia}

\affiliation{$^2$Peter Gr\"unberg Institut,
Forschungszentrum J\"ulich, D-52425, Germany}

\begin{abstract}
We calculate the quantum heat generation, the interaction force  and the frictional 
torque for two rotating spherical nanoparticles with a radius $R$. In contrast to the static 
case, when there is an upper 
limit in the radiative heat transfer between the particles, for two rotating nanoparticles the 
quantum  heat generation rate diverges when 
the angular velocity becomes equal   to  the poles in the photon emission rate.  
These   poles arise  
for the separation $d <d_0= R(3/\varepsilon''(\omega_0))^{1/3}$ (where $\varepsilon''(\omega_0)$
is the imaginary part of the dielectric function for the particle material at the surface 
phonon or plasmon polariton frequency  
 $\omega_0$ ) due to the anomalous Doppler 
effect and  the mutual  polarization of 
the particles and they exist   even  for the particles with losses. Similar singularities exist 
also for the interaction force and the frictional torque. The obtained results 
can be important for biomedical applications. 
\end{abstract}

\maketitle


\vskip 5mm

\section{Introduction}

At present a great deal of attention is devoted to the study of rotating nanoparticles 
in the context of wide variety of physical, chemical and biomedical applications. 
The most important  are related to using of rotating nanoparticles for targeting of 
cancer cells \cite{Nanomed2014,TrendsBioTech2011,ACSNano2014}. The frictional forces 
due to quantum fluctuations acting on a small sphere rotating near a
surface were studied in Ref. \cite{PendryPRL2012,DedkovEPL2012}. Different experimental 
setups for trapping and rotating nanoparticles were discussed recently 
in Refs. \cite{nacommun2011,nanano2013,nanolett2014}. 

Two arbitrary media in relative motion or at rest and separated by a vacuum gap 
continually exchange  energy and momentum via a fluctuating electromagnetic field 
which is always present in the vacuum gap due to thermal and quantum fluctuations 
inside media \cite{VolokitinRMP2007}. This energy and momentum transfer is responsible 
for the radiative heat transfer and non-contact friction. At the nanoscale, these 
phenomena are enhanced by many orders of  magnitude due to the contribution 
from  evanescent electromagnetic waves. Further enhancement occurs if the media can 
support surface phonon- or plasmon- polariton modes. The possibility of using localized 
photon tunnelling between adsorbate vibrational modes for heating 
of the molecules was discussed in Ref.\cite{PerssonPRB2007}. All these phenomena raised 
a fundamental question. Are there limits which restrict the efficiently of  energy 
and momentum transfer betweens bodies? For the static case in the far-field the radiative 
heat transfer is maximal for blackbodies when it is described by the Stefan-Boltzmann 
law. In the near-field the upper limit for the radiative heat transfer is determined by 
 the transmission coefficient for photon tunnelling 
which can not exceed unity in the static case
\cite{PendryJPCM1997,VolokitinJETPLett2003,JoulainPRB2010}. However for two sliding 
plates the photon emission rate  can diverge at the resonant conditions due to the 
anomalous  Doppler effect \cite{JacobJOpt2014,JacobOptExp2014,VolokitinPRB2016}. 

In this article we calculate the frictional 
torque,  the interaction force and the heat generation for two rotating nanoparticles
using  fluctuation electrodynamics. 
We determine the resonance conditions under which these quantities have singularities 
due to  the mutual polarization  of
the particles and the anomalous Doppler effect.

\begin{figure}
\includegraphics[width=0.60\textwidth]{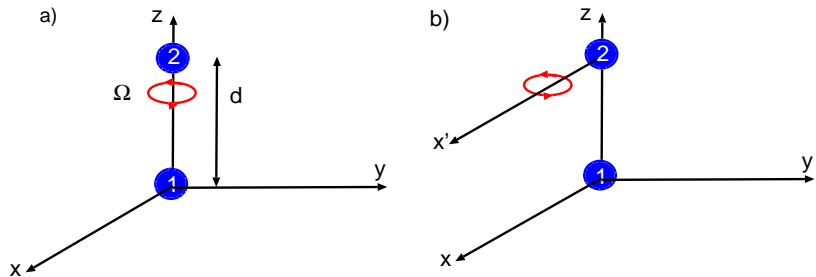}
\caption{A nanoparticle 2 rotating along   the $\hat{z}$-axis (Fig.\ref{Scheme}a) 
 and  the $\hat{x{^\prime}}$-axis (Fig.\ref{Scheme}b), and 
located  at a separation 
$d$ from the other nanoparticle 1 at  the origin. \label{Scheme}}
\end{figure}

\section{Theory}

We consider two spherical particles 1 and 2 located along the $\hat{z}$-axis at 
$\textbf{r}_1=(0,0,0)$ and $\textbf{r}_2=(0,0,d)$ (see Fig. \ref{Scheme}). 
They are characterized by the different temperatures $T_1$, $T_2$ and have the frequency-dependent 
polarizabilities $\alpha_{1,2}(\omega)$. 
We introduce two reference frames $K$ and $K^{\prime}$. In the $K$ frame the particle 1 is 
at  rest while the particle 2 rotates around  the  axis passing through it with an angular 
velocity $\Omega$. The $K^{\prime}$ frame is the rest reference frame for the particle 2.
The orientation of the rotation axis for the particle 2 can be arbitrary but in the present 
study  we consider the most symmetric cases when the rotation axis is along $\hat{z}$ 
or $\hat{x}^{\prime}$ axes as on Figures \ref{Scheme}a and \ref{Scheme}b, respectively.
In the comparison with the general case for these limiting cases the calculations are 
much simpler and the obtained results are 
qualitatively the same.

\subsection{Rotation axis  along the $\hat{z}$ axis (see Fig. \ref{Scheme}a)}

According to  fluctuation electrodynamics\cite{VolokitinRMP2007}, the dipole moment for a 
polarizable particle $\mathbf{p}=\mathbf{p}^{f}+\mathbf{p}^{ind}$ 
where $\mathbf{p}_{i}^{f}$ is the fluctuating dipole moment due to  
quantum and thermal fluctuations inside the particle,  $\mathbf{p}_{i}^{ind}$ is the induced 
dipole moment. In the $K$ frame the Fourier 
transformation is determined by
\begin{equation}
\mathbf{p}_{i}(t)=\int_{-\infty}^{\infty}\frac{d\omega}{2\pi}\mathbf{p}_{i}(\omega)
 e^{-i\omega t}.
\label{FT}
\end{equation}  
In the $K$ frame the dipole moment for   particle 1   
satisfies   equation

\begin{equation}
\mathbf{p}_{1}(\omega)=\alpha_1(\omega)\mathbf{E}_{12}(\omega) + 
\mathbf{p}^f_{1}(\omega),
\label{dip1perp}
\end{equation}
where  the electric field created by 
 particle 2 at the position of  particle 1 $\mathbf{E}_{12}$ is given by 
\begin{equation}
\mathbf{E}_{12}(\omega)=\frac{3p_{2z}(\omega)\hat{z}}{d^3}
-\frac{\mathbf{p}_{2}(\omega)}{d^3},
\label{Eperp}
\end{equation}
where the first and second
 terms in the right side of Eq. (\ref{dip1perp})  determine 
the induced and fluctuating 
dipole moments of  particle 1, respectively, $\alpha_1(\omega)$ is 
the polarizability for the particle 1. In the $K^{\prime}$  frame the components 
of the dipole moments $\mathbf{p}_i^{\prime}$  satisfy to the equation similar to the  
Eq.  
(\ref{dip1perp})
\begin{equation}
\mathbf{p}_{2}^{\prime}(\omega)=\alpha_2(\omega)\mathbf{E}^{\prime}_{21}(\omega) + 
\mathbf{p}_{2}^{f\prime}(\omega),
\label{dip2perpprime}
\end{equation}  
where $\mathbf{E}^{\prime}_{21}$ is the electric field created by the particle 1 at the 
position of the particle 2. The relations between the dipole moment of the particle 2 in 
the $K$ and 
$K^\prime$ frames are determined 
by the equations: $p_{2z}^{\prime}(t)=p_{2z}(t)$ and
\begin{equation}
\mathbf{p}^{\prime}_{2\perp}(t)=
\left(
\begin{array}{cc}
\cos \Omega t& \sin \Omega t\\
-\sin \Omega t &\cos \Omega t
\end{array}\right)\mathbf{p}_{i\perp}(t),
\label{T}
\end{equation}
where $\mathbf{p}^{\prime}_{2\perp}=(p_x, p_y)$, and for the Fourier components: 
$p_{2z}^{\prime}(\omega)=
p_{iz}(\omega)$,    
\begin{equation}
\mathbf{p}^{\prime}_{2\perp}
(\omega)=
\hat{e}^{\prime +}p^-_2({\omega^+}) +
\hat{e}^{\prime -}p_2^+({\omega^-})
\label{FTdip}
\end{equation}
where $\omega^{\pm}=\omega \pm \Omega$, $\hat{e}^{\prime \pm}=(\hat{x}^{\prime}\pm 
i\hat{y}^{\prime})/\sqrt{2}$, $p_2^{\pm}=(p_{2x}\pm 
ip_{2y})/\sqrt{2}$. The same relations are valid  for the $\mathbf{E}^{\prime}_{21}$: 
$E_{21z}^{\prime}(\omega)=
E_{21z}(\omega)$,
\begin{equation}
\mathbf{E}^{\prime}_{21\perp}
(\omega)=
\hat{e}^{\prime +}E^-_{21}({\omega^+}) +
\hat{e}^{\prime -}E_{21}^+({\omega^-}),
\label{FTField}
\end{equation}
where $E_{21}^{\pm}=(E_{21x}\pm 
iE_{21y})/\sqrt{2}$,
\begin{equation}
\mathbf{E}_{21}(\omega)=\frac{3p_{1z}(\omega)\hat{z}}{d^3}
-\frac{\mathbf{p}_{1}(\omega)}{d^3},
\label{2Eperp}
\end{equation}
Using these relations in Eq. (\ref{dip2perpprime}) and taking into account that
$(\hat{e}^{\pm}\cdot \hat{e}^{\mp})=1$, $(\hat{e}^{\pm}\cdot \hat{e}^{\pm})=0$ and  
$(\hat{e}_z\cdot \hat{e}^{\pm})=(\hat{e}^{\pm}\cdot \hat{e}_z)=0$ 
  we get
\begin{equation}
p_{2z}(\omega)=\frac{2\alpha_2(\omega)p_{1z}(\omega)}{d^3}+p_{2z}^f(\omega),
\label{p2z1z}
\end{equation}
\begin{equation}
p_{2x}(\omega)+ip_{2y}(\omega)=-\frac{\alpha_2(\omega^+)
[p_{1x}(\omega)+ip_{1y}(\omega)]}{d^3}
+p_{2}^{f\prime +}(\omega^+),
\label{p2+p1+}
\end{equation}
\begin{equation}
p_{2x}(\omega)-ip_{2y}(\omega)=-\frac{\alpha_2(\omega^-)
[p_{1x}(\omega)-ip_{1y}(\omega)]}{d^3}
+p_{2}^{f\prime -}(\omega^-),
\label{p2-p1-}
\end{equation}
where $p_2^{f\pm}(\omega^\pm)=
p_{2x}^{f\prime}(\omega^\pm)\pm ip_{2y}^{f\prime}(\omega^\pm)$.
From Eqs. (\ref{dip1perp}) and (\ref{p2z1z})-(\ref{p2-p1-}) we get:
\begin{equation}
p_{1z}(\omega)=\frac{p_{1z}^f(\omega) + 2\alpha_1(\omega)p_{2z}^f(\omega)/d^3}
{1-4\alpha_1(\omega)\alpha_2(\omega)/d^6},
\label{p1z}
\end{equation}
\begin{equation}
p_{2z}(\omega)=\frac{p_{2z}^f(\omega) + 2\alpha_2(\omega)p^{f}_{1z}/d^3}{1-
4\alpha_1(\omega)\alpha_2(\omega)/d^6},
\label{p2z}
\end{equation}
\begin{equation}
\mathbf{p}_{1x}(\omega)=\frac 1{2}\left[\frac{p_1^{f+}(\omega)  
-\alpha_1(\omega)
p_{2}^{f\prime +}(\omega^+)/d^3}
{D^+}+\frac{p_1^{f-}(\omega)  
-\alpha_1(\omega)
p_{2}^{f\prime -}(\omega^-)/d^3}
{D^-}\right],
\label{p1x}
\end{equation}
\begin{equation}
\mathbf{p}_{1y}(\omega)=\frac 1{2i}\left[\frac{p_1^{f+}(\omega)  
-\alpha_1(\omega)
p_{2}^{f\prime +}(\omega^+)/d^3}
{D^+}-\frac{p_1^{f-}(\omega)  
-\alpha_1(\omega)
p_{2}^{f\prime -}(\omega^-)/d^3}
{D^-}\right],
\label{p1y}
\end{equation}
\begin{equation}
\mathbf{p}_{2x}(\omega)=\frac 1{2}\left[\frac{p_2^{f\prime+}(\omega^+)  
-\alpha_2(\omega^+)
p_{1}^{f +}(\omega)/d^3}
{D^+}+\frac{p_2^{f\prime-}(\omega^-)  
-\alpha_2(\omega^-)
p_{1}^{f -}(\omega)/d^3}
{D^-}\right],
\label{p2x}
\end{equation}
\begin{equation}
\mathbf{p}_{2y}(\omega)=\frac 1{2i}\left[\frac{p_2^{f\prime+}(\omega^+)  
-\alpha_2(\omega^+)
p_{1}^{f +}(\omega)/d^3}
{D^+}-\frac{p_2^{f\prime-}(\omega^-)  
-\alpha_2(\omega^-)
p_{1}^{f -}(\omega)/d^3}
{D^-}\right],
\label{p2x}
\end{equation}
where $D^{\pm}=1-\alpha_1(\omega)\alpha_2(\omega^{\pm})/d^6$, $p_1^{f\pm}(\omega)=
p_{1x}^{f}(\omega)\pm ip_{1y}^{f}(\omega)$. The spectral density of the
 fluctuations of   the dipole 
moment of the $i$-th particle in the rest reference frame of the  particle is 
determined 
by the fluctuation dissipation theorem   
\begin{equation}
<p^{f}_{ij}(\omega)p^{f*}_{ik}(\omega^{\prime})>=2\pi \delta(\omega-\omega^{\prime})
<p^{f}_{ij}p^{f}_{ik})>_{\omega}
\label{FDT}
\end{equation}
where
\begin{equation}
<p^{f}_{ij}p^{f}_{ik}>_{\omega}=\hbar\mathrm{Im}\alpha_i(\omega)
\mathrm{coth}\left(\frac{\hbar\omega}{2k_BT_i}\right)\delta_{jk}.
\end{equation}
The torque acting on the particle 1 along the $\hat{z}$-axis can be written 
in the form
\begin{equation}
M_z=\int_{-\infty}^{\infty}\frac{d\omega}{2\pi}
<[p_{1x}E_{12y}-p_{1y}E_{12x}]>_{\omega}
\label{Torque1}
\end{equation}
where $\mathbf{E}_{12}$
is the electric field created by the particle 2 at the position 
of the particle 1.
Using Eqs. (\ref{p1z})-(\ref{FDT}) we get
\begin{equation}
M_z=\frac{\hbar}{\pi d^6}\int_{-\infty}^{\infty}d\omega\frac{\mathrm{Im}\alpha_1(\omega)
\mathrm{Im}\alpha_2(\omega^-)}{|1-\alpha_1(\omega)\alpha_2(\omega^-)/d^6|^2}
\left(\mathrm{coth}\frac{\hbar \omega^-}{2k_BT_2}-\mathrm{coth}
\frac{\hbar \omega}{2k_BT_1}\right)
\label{Torque2}
\end{equation}
The contribution to the torque from the quantum fluctuations (quantum friction) 
which exists even for $T_1=T_2=0$K
\begin{equation}
M_{zQ}=-\frac{2\hbar}{\pi d^6}\int_0^{\Omega}d\omega\frac{\mathrm{Im}\alpha_1(\omega)
\mathrm{Im}\alpha_2(\omega^-)}{|1-\alpha_1(\omega)\alpha_2(\omega^-)/d^6|^2}
\label{TorqueQ}
\end{equation}
The heat generated in the particle 1 by a fluctuating electromagnetic field is 
determined by
\[
P_1 = \int_{-\infty}^{\infty}\frac{d\omega}{2\pi}
<\mathbf{j}_1\cdot \mathbf{E}_{12}>_{\omega}
=\int_{-\infty}^{\infty}\frac{d\omega}{2\pi}
<-i\omega\mathbf{p}_1\cdot \mathbf{E}_{12}>_{\omega}
\]
\[
 = \frac{\hbar}{\pi d^6}\int_{-\infty}^{\infty}d\omega\omega
\left[2\frac{\mathrm{Im}\alpha_1(\omega)
\mathrm{Im}\alpha_2(\omega)}{|1-4\alpha_1(\omega)\alpha_2(\omega)/d^6|^2}
\left(\mathrm{coth}\frac{\hbar \omega}{2k_BT_2}-
\mathrm{coth}\frac{\hbar \omega}{2k_BT_1}\right) \right.
\]

\begin{equation}
\left. +\frac{\mathrm{Im}\alpha_1(\omega)
\mathrm{Im}\alpha_2(\omega^-)}{|1-\alpha_1(\omega)\alpha_2(\omega^-)/d^6|^2}
\left(\mathrm{coth}\frac{\hbar \omega^-}{2k_BT_2}-\mathrm{coth}
\frac{\hbar \omega}{2k_BT_1}\right)\right]
\label{Q1}
\end{equation}
and the heat generated by the quantum fluctuations is given by
\begin{equation}
P_{1Q}=-\frac{2\hbar}{\pi d^6}\int_{0}^{\Omega}d\omega\omega\frac{\mathrm{Im}\alpha_1(\omega)
\mathrm{Im}\alpha_2(\omega^-)}{|1-\alpha_1(\omega)\alpha_2(\omega^-)/d^6|^2}.
\label{P1Q}
\end{equation}
The force acting  on the particle 1 along the $\hat{z}$-axis  is 
given by
\[
F_{1z} =\int_{-\infty}^{\infty}\frac{d\omega}{2\pi}
<\mathbf{p}_1\cdot \frac{d}{dz}\mathbf{E}_{12}(z\rightarrow 0)>_{\omega}
\]
\[
 = \frac{\hbar}{\pi d^7}\int_{-\infty}^{\infty}d\omega
\left[\frac{6}{|1-4\alpha_1(\omega)\alpha_2
(\omega)/d^6|^2}\left(\mathrm{Im}\alpha_1(\omega)
\mathrm{Re}\alpha_2(\omega)\mathrm{coth}\frac{\hbar \omega}
{2k_BT_1}+\mathrm{Re}\alpha_1(\omega)
\mathrm{Im}\alpha_2(\omega)\mathrm{coth}\frac{\hbar \omega}{2k_BT_2}\right)
 \right.
\]
\begin{equation}
\left. +\frac{3}{|1-\alpha_1(\omega)\alpha_2(\omega^-)/d^6|^2}
\left(\mathrm{Re}\alpha_1(\omega)
\mathrm{Im}\alpha_2(\omega^-)\mathrm{coth}\frac{\hbar \omega^-}{2k_BT_2}
+\mathrm{Im}\alpha_1(\omega)
\mathrm{Re}\alpha_2(\omega^-)\mathrm{coth}
\frac{\hbar \omega}{2k_BT_1}\right)\right]
\label{F1}
\end{equation}
The contribution to $F_{1z}$ from the frequency region 
corresponding to the anomalous Doppler effect 
in Eq.(\ref{F1}) is determined by the integration in the interval 
$0<\omega<\Omega$ and for $T_1=T_2=0$K is given by
\begin{equation}
F_{1z}^{AD}= \frac{\hbar}{\pi d^7}\int_{0}^{\Omega}d\omega
\frac{3}{|1-\alpha_1(\omega)\alpha_2(\omega^-)/d^6|^2}
\left[
\mathrm{Im}\alpha_1(\omega)
\mathrm{Re}\alpha_2(\omega^-)-\mathrm{Re}\alpha_1(\omega)
\mathrm{Im}\alpha_2(\omega^-)\right].
\label{FAD}
\end{equation}

\subsection{The rotation axis is along the $\hat{x}^{\prime}$ axis (see Fig. \ref{Scheme}b)
\label{Bconf}}

The details of the calculations for the case when the rotation axis is along 
the $\hat{x}^{\prime}$ axis  are given in Appendix \ref{A}. 
These calculations are more involved in comparison with the case when 
the rotation axis is along 
the $\hat{z}$ axis.  Using Eqs. (\ref{p1zx})-(\ref{p2zx}) we get 
the resulting  formulas  for $M_x,\,P_1,\, F_{1z}$ and  $F_{1y}$:
\[
M_x=\int_{-\infty}^{\infty}\frac{d\omega}{2\pi}
<[p_{1y}E_{12z}-p_{1z}E_{12y}]>_{\omega}
\]
\begin{equation}
=\frac{8\hbar}{\pi d^6}\int_{-\infty}^{\infty}d\omega
\frac{\mathrm{Re}(D_1^{+*}D_2^{+})\mathrm{Im}\alpha_1(\omega)
\mathrm{Im}\alpha_2(\omega^-)}{|\Delta|^2}
\left(\mathrm{coth}\frac{\hbar \omega^-}{2k_BT_2}-\mathrm{coth}
\frac{\hbar \omega}{2k_BT_1}\right),
\label{Mx}
\end{equation}
\[
 P_1= \frac{\hbar}{2\pi d^6}\int_{-\infty}^{\infty}d\omega\omega
\left[\frac{\mathrm{Im}\alpha_1(\omega)
\mathrm{Im}\alpha_2(\omega)}{|1-\alpha_1(\omega)\alpha_2(\omega)/d^6|^2}
\left(\mathrm{coth}\frac{\hbar \omega}{2k_BT_2}-
\mathrm{coth}\frac{\hbar \omega}{2k_BT_1}\right) \right.
\]

\begin{equation}
\left. +\frac{4(|D_1^+|^2+4|D_2^+|^2)\mathrm{Im}\alpha_1(\omega)
\mathrm{Im}\alpha_2(\omega^-)}{|\Delta|^2}
\left(\mathrm{coth}\frac{\hbar \omega^-}{2k_BT_2}-\mathrm{coth}
\frac{\hbar \omega}{2k_BT_1}\right)\right]
\label{Q1}
\end{equation}
\[
 F_{1z} = \frac{\hbar}{2\pi d^7}\int_{-\infty}^{\infty}d\omega
\left[\frac{3}{|1-\alpha_1(\omega)\alpha_2(\omega)/d^6|^2}
\left(\mathrm{Re}\alpha_1(\omega)
\mathrm{Im}\alpha_2(\omega)\mathrm{coth}\frac{\hbar \omega}{2k_BT_2}+
\mathrm{Im}\alpha_1(\omega)
\mathrm{Re}\alpha_2(\omega)\mathrm{coth}\frac{\hbar \omega}{2k_BT_1}\right) \right.
\]

\begin{equation}
\left. +\frac{12(|D_1^+|^2+4|D_2^+|^2)}{|\Delta|^2}
\left(\mathrm{Re}\alpha_1(\omega)
\mathrm{Im}\alpha_2(\omega^-)\mathrm{coth}\frac{\hbar \omega^-}{2k_BT_2}+
\mathrm{Im}\alpha_1(\omega)
\mathrm{Re}\alpha_2(\omega^-)\mathrm{coth}
\frac{\hbar \omega}{2k_BT_1}\right)\right]
\label{Fzx}
\end{equation}
\[
F_{1y} =\int_{-\infty}^{\infty}\frac{d\omega}{2\pi}
<\mathbf{p}_1\cdot \frac{d}{dy}\mathbf{E}_{12}(y\rightarrow 0)>_{\omega}
=\frac{3}{d^4}\int_{-\infty}^{\infty}\frac{d\omega}{2\pi}<p_{1y}
p_{2z}+p_{1z}
p_{2y}>_{\omega}
\]
\[
 = \frac{\hbar}{\pi d^7}\int_{-\infty}^{\infty}d\omega \frac{6}{|\Delta|^2}
\left[3\mathrm{Re}(D_1^+D_2^{+*})\mathrm{Im}\alpha_1(\omega)\mathrm{Im}\alpha_2(\omega^-)
\left(\mathrm{coth}\frac{\hbar \omega^-}{2k_BT_2}-
\mathrm{coth}\frac{\hbar \omega}{2k_BT_1}\right) \right.
\]
\begin{equation}
\left. -\mathrm{Im}(D_1^+D_2^{+*})
\left(\mathrm{Re}\alpha_1(\omega)
\mathrm{Im}\alpha_2(\omega^-)\mathrm{coth}\frac{\hbar \omega^-}{2k_BT_2}+
\mathrm{Im}\alpha_1(\omega)
\mathrm{Re}\alpha_2(\omega^-)\mathrm{coth}
\frac{\hbar \omega}{2k_BT_1}\right)\right],
\label{Fyx}
\end{equation}
where $D_1^{\pm}=1-4\alpha_1(\omega)\alpha_2(\omega^{\pm})/d^6$, 
$D_2^{\pm}=1-\alpha_1(\omega)\alpha_2(\omega^{\pm})/d^6$ and 
$\Delta=D^+_1D^-_2+D^-_1D^+_2$.

\section{Resonant heat transfer and heat generation due to quantum friction}

For $\Omega =0$ the transmission coefficient for the photon tunnelling for two 
identical particles is restricted by 
the condition\cite{VolokitinRMP2007,VolokitinJETPLett2003}
\begin{equation}
t^T=\frac {4(\mathrm{Im}\alpha/d^3)^2}{|1-(\alpha/d^3)^2|^2} \leq 1.
\end{equation}
Thus $P\leq P_{max}$ where
\begin{equation}
P_{max}=\frac{\pi k_B^2}{2\hbar}\left(T_2^2 - T_1^2\right).
\end{equation}

The radiative heat transfer between two particles is strongly enhanced 
in the case of the resonant photon tunnelling 
\cite{VolokitinRMP2007,VolokitinJETPLett2003}.
For a spherical 
particle of radius $R$ the particle polarizability is given by
\begin{equation}
\alpha_i(\omega) = R^3\frac{\varepsilon_i -1}{\varepsilon_i +2}
\label{polarizability}
\end{equation}
where $\varepsilon_i$ is the dielectric function for a material of sphere. A particle 
has the resonance at $\varepsilon^{\prime}(\omega_i)=-2$ where $\varepsilon^{\prime}$ is the 
real part of $\varepsilon$. For a polar dielectric $\omega_i$ determines the frequency
of the surface phonon-polariton. Close to the resonance for $\omega \approx 
\omega_i$ the particle polarizability 
can be written in the form
\begin{equation}
\alpha_i (\omega) \approx -R^3\frac{a_i}{\omega - \omega_i +i\Gamma_i}
\label{res}
\end{equation}
where
\begin{equation}
a_i=\frac {3}{(d/d\omega)\varepsilon_i^{\prime}(\omega)|_{\omega=\omega_i}},\,\,\,
\Gamma = \frac {\mathrm{Im}\varepsilon_i(\omega_i)}{(d/d\omega)
\varepsilon_i^{\prime}(\omega)|_{\omega=\omega_i}}
\end{equation}
Close to the resonance for two identical particles ($\omega_1=\omega_2=\omega_0$, 
$a_1=a_2=a$) the transmission coefficient 
can be written in the form
\begin{equation}
t^T\approx \frac {4[a\Gamma(R/d)^3]^2}{[(\omega-\omega_+)^2+\Gamma^2]
[(\omega-\omega_-)^2+\Gamma^2]}
\label{tapp}
\end{equation}
where $\omega_\pm = \omega_0 \pm a(R/d)^3 $. For $a(R/d)^3>\Gamma$ the resonant 
heat transfer is given by
\begin{equation}
P_{res}\approx 6\hbar \omega_0\Gamma[n_1(\omega_0)-n_2(\omega_0)]
\label{Pres}
\end{equation}
where $n_i(\omega)=[\exp(\hbar \omega/k_BT_i)-1]^{-1}$. For $\hbar\omega_0 <
k_BT_i$ $P_{res}\approx 6\Gamma k_B(T_2-T_1)$ and for $T_2\gg T_1$
\begin{equation}
\frac{P_{res}}{P_{max}}\approx \frac{12}{\pi}\left(\frac{\hbar \Gamma}{k_BT_2}
\right)
< \left(\frac{\hbar \omega_0}{k_BT_2}\right)< 1.
\end{equation}
For $a(R/d)^3<\Gamma$
\begin{equation}
P_{res}\approx \frac{\hbar \omega_0 a^2}{\Gamma}\left(\frac{R}{d}\right)^6
[n_2(\omega_0)-n_1(\omega_0)]<\hbar \omega_0\Gamma[n_2(\omega_0)-n_1(\omega_0)]. 
\end{equation}

Another resonance is possible in the condition of the anomalous Doppler 
effect when $\omega_1-\Omega=-\omega_2$ \cite{VolokitinRMP2007,VolokitinPRL2011,
VolokitinEPL2013,JacobJOpt2014,JacobOptExp2014}. 
At this resonant condition, taking into account that
\begin{equation} 
\alpha_1(\omega_1)\approx e^{\frac {i\pi}{2}}|\alpha_1(\omega_1)|,  \,\,
\alpha_2(-\omega_2)\approx e^{-\frac{i\pi}{2}}|\alpha_2(\omega_2)| 
\end{equation}
the denominators in the 
integrands in Eqs. (\ref{TorqueQ}), (\ref{P1Q}) and (\ref{FAD}) contain the factor
\begin{equation}
1-\frac{|\alpha_1(\omega_1)\alpha_2(\omega_2)|}{d^6}.
\end{equation}
At the resonance $|\alpha_1(\omega_1)\alpha_2(\omega_2)|/d^6$ can be larger 
than unity thus the denominator 
is equal to zero at 
\begin{equation}
d_0=(|\alpha_1(\omega_1)\alpha_2(\omega_2)|^{1/3}
\label{dc}
\end{equation}
what means that for $d<d_0$ the friction torque, heat generation and force 
interaction can diverge. 
The origin of this divergence is 
related to the creation below critical separation   $d_0$  of the  
resonance at frequency determined by the pool of the photon 
emission rate   for two rotating particle. This resonance can be  lossless  even in the case 
when the surface phonon- polariton modes for the isolated particles have losses. 
At such  
critical conditions the amplitude of electric field increases infinitely 
with time which gives rise to the divergence of the heat generation  
and interaction forces \cite{SilveirinhaNJP2014}.

Substituting  Eq.(\ref{res}) in  Eq.(\ref{dc}) for the critical separation 
we get  
\begin{equation}
d_0=R\left(\frac{a_1a_2}{\Gamma_1\Gamma_2}\right)^{1/6}=
R\left(\frac{9}{\varepsilon_1''(\omega_1)\varepsilon_2''(\omega_2)}\right)^{1/6},
\end{equation}
(for example, for silicon carbide (SiC) $d_0= 2.57R$ (see below)) and the 
polarizabilities 
for the particles 1 and 2 for $\omega\approx \omega_1$ and $\omega-
\Omega\approx\omega_2$ are 
given by Eq. (\ref{res})  
and by equation
\begin{equation}
\alpha_2(\omega-\Omega)\approx -R^3\frac{a_2}{ 
\Omega-\omega_2-\omega -i\Gamma_2},
\label{antires}
\end{equation}
respectively. In this resonant case the photon emission rate for $0<\omega<\Omega$ is 
given by equation
\[
t^E=\frac{4\mathrm{Im}\alpha_1(\omega)
\mathrm{Im}\alpha_2(\omega-\Omega)/d^6}
{|1-\alpha_1(\omega)\alpha_2(\omega-\Omega))/d^6|^2}
\]
\begin{equation}
\approx
\frac{4\Gamma_1\Gamma_2a_1a_2(R/d)^6}
{(\Gamma_1+\Gamma_2)^2(\omega -\omega_c)^2
+\left[\Gamma_1\Gamma_2\left(\frac{\Omega-\Omega_0}{\Gamma_1+\Gamma_2}
\right)^2
-(\omega -\omega_c)^2+ \frac{(\Omega-\Omega_0)(\Gamma_2-\Gamma_1)(\omega-\omega_c)}
{\Gamma_1+\Gamma_2}+
\Gamma_1\Gamma_2-a_1a_2(R/d)^6\right]^2}
\label{tantipart}
\end{equation}
where $\Omega_0=\omega_1+\omega_2$,
\begin{equation}
\omega_c=\frac{\Gamma_1(\Omega-\omega_2)+\Gamma_2\omega_1}{\Gamma_1+\Gamma_2}
\end{equation}

For two identical particles the transmission coefficient  diverges at 
 $\omega=\omega_c=\omega_0$ and 
$\Omega =
\Omega^{\pm}$ where
\begin{equation}
\Omega^{\pm}=2\left[\omega_0\pm\Gamma\sqrt{\left(\frac{a}{\Gamma}\right)^2
\left(\frac{R}{d}\right)^6-1}\right].
\end{equation}
Close to the resonance when
\begin{equation}
\frac{1}{4}\left|\left(\frac{\Omega - \Omega_0}{2\Gamma}\right)^2
+1-\left(\frac{a}{\Gamma}\right)^2\left(\frac{R}{d}\right)^6\right|\ll 1
\end{equation}
 using Eq. (\ref{tantipart}) in Eq. (\ref{P1Q})  we get
\begin{equation}
P_{1Q}\approx \frac{\hbar \omega_0}{\Gamma}\frac{a^2(R/d)^6}
{\left|\left(\frac{\Omega - \Omega_0}{2\Gamma}\right)^2
+1-\left(\frac{a}{\Gamma}\right)^2\left(\frac{R}{d}\right)^6\right|}.
\label{P1Q2}
\end{equation}
At $\Omega=\Omega_0$ the photon emission rate diverges at $\omega=\omega_1$ 
and $d=d_0$. Close to this resonance quantum heat generation behaves  as
\begin{equation}
P_{1Q}\propto \frac{d_0}{|d-d_0|}
\end{equation}

\begin{figure}
\includegraphics[width=0.80\textwidth]{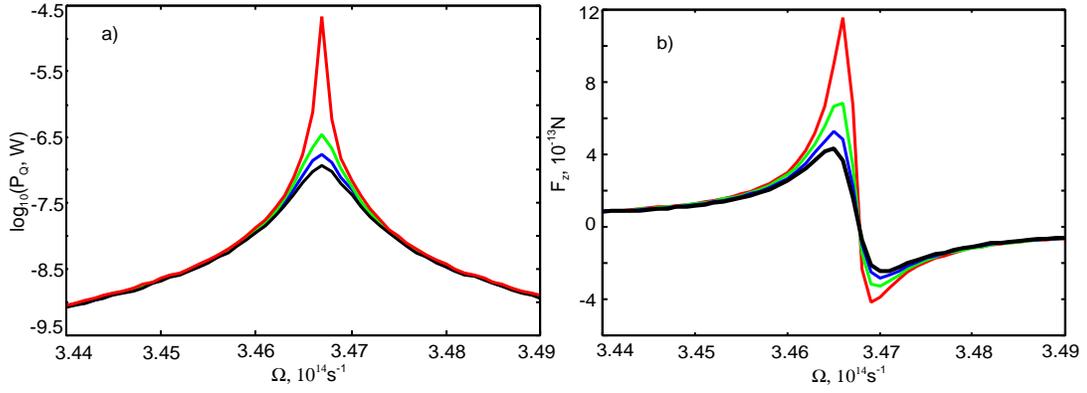}
\caption{ a) The  dependence of the heat generation rate due to quantum friction  
for the SiC particle 1 
with a radius R =
0.5nm, and b) the interaction forces   between the particles on the rotation frequency 
$\Omega$  of the same particle 2. The  red,
green, blue and black lines  show the results of the calculations  
 for $d>d_0=2.57R$ at  $d=2.60R$,  $d=2.61R$,  $d=2.62R$
and  $d=2.63R$, respectively, where  $d_0$ is the critical separation  between the particles, 
below which    the quantum heat  generation  rate diverges at the resonant frequencies 
$\Omega^{\pm}$.
\label{SQ}}
\end{figure}

\begin{figure}
\includegraphics[width=0.80\textwidth]{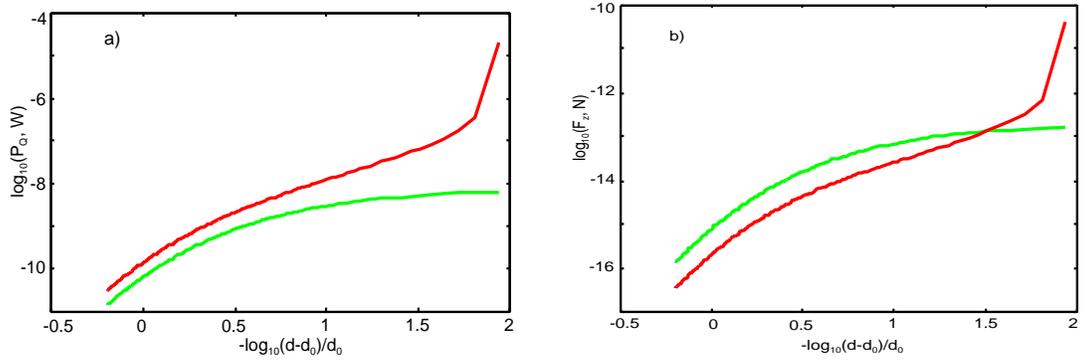}
\caption{a) The  dependence of the heat generation rate due to quantum friction  
for the SiC particle 1 
with a radius R =
0.5nm, and b) the interaction force   on the separation  between the particles.
The  red and 
green lines  show the results of the calculations for $\Omega=\Omega_0=2\omega_0$ and 
$\Omega=\Omega_0(1+0.003)$, respectively, where $\omega_0$ is the surface phonon polariton 
frequency for a SiC particle. 
\label{FAD}}
\end{figure}

As an example,
consider two nanoparticles of silicon carbide (SiC). The optical
properties of this material can be described using an oscillator
model \cite {Palik}
\begin{equation}
\varepsilon (\omega )=\epsilon _\infty \left( 1+\frac{\omega
_L^2-\omega _T^2}{ \omega _T^2-\omega ^2-i\Gamma \omega }\right),
\label{1nine}
\end{equation}
with $\varepsilon _\infty =6.7$, $\omega _L=1.8\cdot 10^{14}$s$^{-1}$, $\omega
_T=1.49\cdot 10^{14}$s$^{-1}$, and $\Gamma =8.9\cdot 10^{11}$s$^{-1}$. The
frequency of surface phonon polaritons  is determined by the condition 
$\varepsilon^{\prime}(\omega
_0)=-2$ and from (\ref{1nine}) we get $\omega _0=1.73\cdot 10^{14}$s$^{-1}$.
From Eq.(\ref{dc}) we get the critical distance $d_0=2.57R$. 

For a particle rotating around the $x^{\prime}$ axis  
the denominators in the integrands in Eqs. ($\ref{Mx}$)-($\ref{Fyx})$ contain 
the factor $\Delta=D^+_1D^-_2+D^-_1D^+_2$ 
(see Sec.\ref{Bconf}). Under the resonance 
conditions when $\omega \approx \omega_0$ and 
$\omega - \Omega \approx -\omega_0$ we can put $D^+_1\approx D^+_2\approx 1$. Thus, 
a resonance 
occurs when   
\begin{equation}
\Delta \approx 2\left(1-\frac{2.5\alpha_1(\omega)\alpha_2(\omega^{-})}{d^6}\right)=0.
\label{Deltax}
\end{equation}
From this equation we get that for the SiC particles the divergence in the photon emission 
rate occurs for $d<d_0=3R$. For an arbitrary orientation of the rotation axis, the critical 
separation for SiC particles is in the range: $2.57R<d_0<3R$.

Fig.\ref{SQ} shows the dependence of a) the  quantum heat  generation rate  for a  particle 1 
and b) the interaction force between
the particles on  the angular velocity of the particle 2 for $d\geq 2.6R> d_0$. 
In accordance with the above theoretical analysis 
these dependences have sharp resonance for $d\rightarrow d_0$. For static particles at 
$T_2 = 300$K and $T_1 = 0$K
from Eq.(\ref{Pres})  follows  that  the resonant photon 
tunneling contribution to 
the radiative heat transfer $P_{res}\approx 10^{-9}$W. In sharp contrast to the static case, 
for rotating particles the heat generation rate  
 diverges at the resonance at  $d = d_0$ and $\Omega = \Omega_0 = 2\omega_0$. 
At the  resonance the stationary 
rotation of a particle is impossible, since in this case the friction force  
increases unrestrictedly  with time. However, near the 
resonance the stationary rotation with
an arbitrarily high heat generation rate   
due to conversion of the mechanical energy into heat is possible. 
Near the resonance frequency, the interaction force changes sign (see Fig.\ref{SQ}b). 
In the static case, the van der Waals force between
two particles is given by formula
\begin{equation}
F_{vdW}(d)=\frac{32}{3}\left(\frac{R}{d}\right)^6\frac{A_H}{d},
\label{FvdW}
\end{equation}
where according to Ref.\cite{SvetovoyPRB2014} the Hamaker constant for the SiC-SiC system 
$A_H=16.5\cdot10^{-20}$J. For $d = 2.6R = 1.3$ nm
$F_{vdW} = 5.7 \cdot 10^{-12}$N. For rotating particles near resonance 
the interaction force can be arbitrarily
 large. Thus tuning of the interaction force is possible by changing the  angular 
velocity  of a particle.

Fig.\ref{FAD} 
shows the dependences of a)  the heat generation rate  and b) the interaction forces 
between the particles on
the separation  between the particles for $d\geq 2.6R> d_0$ for $\Omega = \Omega_0$
 (red curve) and $\Omega = \Omega_0 (1 + 0.003)$ (green
curve). In accordance with the above theoretical analysis these dependences
have divergences at the critical angular velocity $\Omega_0$.

The condition for the validity of the dipole approximation for two 
particles is determined by 
$2R/d\ll 1$. For SiC particles, the multipole expansion parameter 
for $d \approx 2.6R \approx d_0$ 
is equal to 0.8 and 0.7 for the rotation axis    directed along and perpendicular 
to the z axis, respectively. Therefore
the numerical calculations given above play the role of a qualitative 
estimation of the effect. Its quantitative
description for SiC particles requires consideration of multipole effects.

\section{Summary}

Fluctuation electrodynamics was used to calculate the heat generation, 
the interaction force and the frictional torque for two rotating nanoparticles, 
taking into account the mutual polarization of the particles. In a sharp contrast to 
the static case, all these quantities 
diverge at the resonant conditions even for the case when there are losses in 
the particles.  The origin of these features is related to the divergence of 
the photon emission  rate under the conditions of the anomalous 
Doppler effect. 
The obtained results can found broad
application in nanotechnology. In particular, they can be used for tuning  
of the interaction forces and the heat generation  by changing the angular velocity. 
These processes can be used for targeting  cancer cells.
For practical application of the predicted effects, it is necessary to search for or 
create materials with a low frequency of the plasmon or phonon polaritons and a small 
imaginary part of the dielectric function at this frequency. InSb semiconductor has a 
frequency of the surface plasmon-phonon polaritons in the THz region\cite {Palik}. However, 
the dielectric 
function for this material has a large imaginary part at this frequency, which leads to 
a small value for the critical distance.
On the other hand metamaterials can have a frequency of the plasmon polaritons 
in the GHz region\cite{PendryJPCM1998}.

\section{Acknowledgement}

The study was supported by  
the Russian Foundation for
Basic Research (Grant No. 16-02-00059-a).

\vskip 0.5cm

\appendix
\section{The rotation axis is along the $\hat{x}^{\prime}$ axis (see Fig. \ref{Scheme}b)
\label{A}}

In the case of the rotation axis directed along the $\hat{x}^{\prime}$ axis instead of 
Eqs. (\ref{p2z1z})-(\ref{p2-p1-}) we get 
\begin{equation}
p_{2x}(\omega)=-\frac{\alpha_2(\omega)p_{1x}(\omega)}{d^3}+p_{2x}^f(\omega),
\label{p2x1x}
\end{equation}
\begin{equation}
p_{2z}(\omega)+ip_{2y}(\omega)=\frac{\alpha_2(\omega^+)
[2p_{1z}(\omega)-ip_{1y}(\omega)]}{d^3}
+p_{2}^{f\prime +}(\omega^+),
\label{p2+p1+x}
\end{equation}
\begin{equation}
p_{2z}(\omega)-ip_{2y}(\omega)=\frac{\alpha_2(\omega^-)
[2p_{1z}(\omega)+ip_{1y}(\omega)]}{d^3}
+p_{2}^{f\prime -}(\omega^-),
\label{p2-p1-x}
\end{equation}
Using Eq. (\ref{dip1perp}) in Eqs. (\ref{p2+p1+x}) and (\ref{p2-p1-x}) we get the set of  
equations:
\begin{equation}
D_1^+p_{2z}(\omega)+iD_2^+p_{2y}=P_2^{f+},
\label{p2zp2y-}
\end{equation}
\begin{equation}
D_1^-p_{2z}(\omega)-iD_2^-p_{2y}=P_2^{f-},
\label{p2zp2y+}
\end{equation}
where $D_1^{\pm}=1-4\alpha_1(\omega)\alpha_2(\omega^{\pm})/d^6$, 
$D_2^{\pm}=1-\alpha_1(\omega)\alpha_2(\omega^{\pm})/d^6$, $P_2^{f\pm}=
\alpha_2(\omega^{\pm})[2p_{1z}^f(\omega)\mp ip_{1y}^f(\omega)]/d^3+p_2^{f\prime \pm}
(\omega^{\pm})$. From Eqs. (\ref{dip1perp}), (\ref{p2x1x}), (\ref{p2zp2y-}) and (\ref{p2zp2y+}) 
we get
\begin{equation}
p_{1x}(\omega)=\frac{p_{1x}^f(\omega) - \alpha_1(\omega)p_{2x}^f(\omega)/d^3}
{1-\alpha_1(\omega)\alpha_2(\omega)/d^6},
\label{p1xx}
\end{equation}
\begin{equation}
p_{2x}(\omega)=\frac{p_{2x}^f(\omega) -\alpha_2(\omega)p^{f}_{1x}(\omega)/d^3}{1-
\alpha_1(\omega)\alpha_2(\omega)/d^6},
\label{p2xx}
\end{equation}
\begin{equation}
p_{1z}=\frac{1}{\Delta}\left[D_2^+P_{1z}^{f-}+D_2^-P_{1z}^{f+}\right],\,\,
p_{1y}=\frac{1}{\Delta}\left[D_1^+P_{1y}^{f-}+D_1^-P_{1y}^{f+}\right],
\label{p1zx}
\end{equation}
\begin{equation}
p_{2z}=\frac{1}{\Delta}\left[D_2^+P_{2}^{f-}+D_2^-P_{2}^{f+}\right],\,\,
p_{2y}=\frac{i}{\Delta}\left[D_1^+P_{2}^{f-}-D_1^-P_{2}^{f+}\right],
\label{p2zx}
\end{equation}
where $P_{1z}^{f\pm}=p_{1z}^f(\omega)\mp 
2ip_{1y}^f(\omega)+2\alpha_1(\omega)p_2^{f\prime \pm}
(\omega^{\pm})/d^3$, $P_{1y}^{f\pm}=p^f_{1y}(\omega)\pm ip^f_{1z}(\omega)/2 
\pm i\alpha_1(\omega)p_2^{f\prime \pm}(\omega^{\pm})/d^3$, 
$\Delta=D^+_1D^-_2+D^-_1D^+_2$.

\vskip 0.5cm

$^*$alevolokitin@yandex.ru

\end{document}